\documentclass[conference]{IEEEtran}
\IEEEoverridecommandlockouts
\usepackage{cite}
\usepackage{amsmath,amssymb,amsfonts}
\usepackage{algorithmic}
\usepackage{graphicx}
\usepackage{textcomp}
\usepackage{xcolor}

\usepackage{amsthm}
\usepackage{url}
\usepackage{bbm}
\usepackage{dsfont}
\usepackage{algorithm}
\usepackage{color}
\usepackage{subfigure}
\usepackage{soul}

\def\BibTeX{{\rm B\kern-.05em{\sc i\kern-.025em b}\kern-.08em
    T\kern-.1667em\lower.7ex\hbox{E}\kern-.125emX}}
\begin{document}
\bstctlcite{BSTcontrol}

\title{Joint Sensing, Communication, and Computation Resource Allocation for Cooperative Perception in Fog-Based Vehicular Networks
}

\author{\IEEEauthorblockN{Xinran~Zhang, Zhimin~He, Yaohua~Sun, Shuo~Yuan, Mugen~Peng\\}
\IEEEauthorblockA{\textit{State Key Laboratory of Networking and Switching Technology} \\
\textit{Beijing University of Posts and Telecommunications}\\
Beijing, China \\
\{xrzhang819, hezhimin-sgcr, sunyaohua, yuanshuo, pmg\}@bupt.edu.cn}
}

\maketitle

\begin{abstract}

    To enlarge the perception range and reliability of individual autonomous vehicles, cooperative perception has been received much attention.
    However, considering the high volume of shared messages, limited bandwidth and computation resources in vehicular networks become bottlenecks. In this paper, we investigate how to balance the volume of shared messages and constrained resources in fog-based vehicular networks. To this end, we first characterize sum satisfaction of cooperative perception taking account of its spatial-temporal value and latency performance. Next, the sensing block message, communication resource block, and computation resource are jointly allocated to maximize the sum satisfaction of cooperative perception, while satisfying the maximum latency and sojourn time constraints of vehicles. Owing to its non-convexity, we decouple the original problem into two separate sub-problems and devise corresponding solutions. Simulation results demonstrate that our proposed scheme can effectively boost the sum satisfaction of cooperative perception compared with existing baselines. 
    
\end{abstract}

\begin{IEEEkeywords}
    Fog-based vehicular networks, cooperative perception, joint sensing, communication, and computation resource allocation, multi-agent deep reinforcement learning
\end{IEEEkeywords}

\section{Introduction}

With the rapid advancements of embedded systems, sensors, and artificial intelligence in recent decades, autonomous driving has evolved from an impossible dream into a foreseeable reality, and will eventually reshape future transport system and driving experience\cite{zhang2020mobile}.
As the key enabling technology, cooperative perception\cite{yang2021machinelearningenabled} exploits vehicle-to-everything (V2X) communications to exchange sensory messages among neighboring vehicles and road side units, and offers an eagle view of broader road environment.
In\cite{wang2018deployment}, a remarkable perception performance gain over traditional separate perception is validated through theoretical analysis, regardless of the additional communication and computation overhead.
Nevertheless, when the number of involved vehicles is $15$, the required communication capacity reaches $500$ Mbps\cite{marvasti2020cooperative,boban2018connecteda}, which overwhelms limited communication bandwidth and computation power in vehicular networks (VNETs) and restricts the implementation of cooperative perception.

To balance the exchanging message volume in cooperative perception and the limited resources in VNETs,
some efforts have been made for the joint allocation of sensing, communication, and computation resource.
Higuchi \emph{et al.}\cite{higuchi2019valueanticipating} attempt to reduce the redundancy of cooperative perception messages by selecting more valuable ones to be delivered. Specifically, the value of messages is firstly quantified based on message history and current mobility, and then more valuable messages are transmitted via V2X communications.
In\cite{abdel-aziz2020vehicular}, a quadtree-based compression mechanism is utilized to partition sensory information into different block messages. Then, the vehicular association, resource allocation, and block message selection are jointly optimized with deep reinforcement learning (DRL) to maximize the satisfaction of cooperative perception.
Du \emph{et al.}\cite{du2020cooperative} go a step further by proposing a platoon-based cooperative perception framework, in which the perception scheduling, computation strategies, and communication resource allocation are jointly optimized to minimize the delay of sensing tasks. 

The aforementioned works\cite{higuchi2019valueanticipating,abdel-aziz2020vehicular,du2020cooperative} commonly concentrate on the direct sharing among vehicles, leaving the more general problem of VNET-assisted cooperative perception less investigated. 
By taking advantages of hierarchical computation resources and flexible resource management, fog-based vehicular networks (FVNETs)\cite{zhang2021delayoptimized} provide a promising edge-cloud collaboration framework.
In this paper, we examine how to realize cooperative perception with constrained resources in FVNETs, where multiple vehicles offload individual sensory messages to fog access points (F-APs) or the cloud server for cooperative computation. The main contribution are summarized three-folds.

\begin{itemize}
    \item Taking spatial-temporal values of sensory messages and latency performance into consideration, we propose a joint sensing block message, communication and computation resource allocation problem to maximize the sum satisfaction of cooperative perception, while satisfying the maximum tolerant latency and sojourn time constraints.
    \item The non-convex sum satisfaction maximization problem is decoupled into a joint block message and offloading mode selection problem plus a joint communication and computation resource allocation problem. By solving two problems based on multi-agent DRL and swap matching, the sum satisfaction can be maximized.
    \item Through simulation results, we show that our proposed scheme can significantly improve the sum satisfaction of cooperative perception in FVNETs.
\end{itemize}

\section{System Model and Problem Formulation}

\begin{figure}[t]
    \centering
    \includegraphics[width=3.4in]{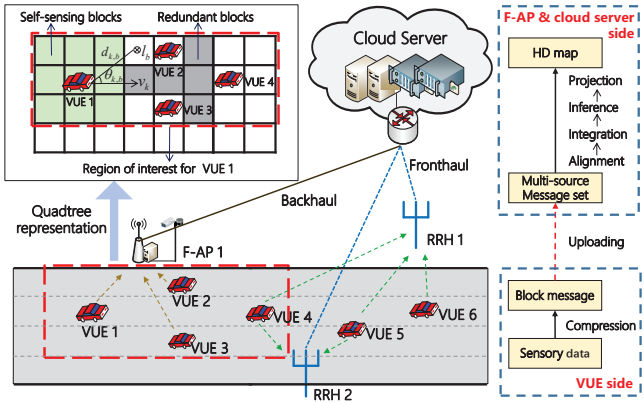}
    \caption{Cooperative Perception in FVNETs.}
    \label{fig:system}
\end{figure}

Considering a cooperative perception scenario in FVNETs, which consists of $K$ vehicular user equipments (VUEs), $N$ F-APs, and $M$ remote radio heads (RRHs) connecting to a centralized cloud server.
As shown in Fig.\ref{fig:system}, each VUE adopts region quadtree technique \cite{abdel-aziz2020vehicular} to compress its sensory data into independent block messages $b \in {{\cal B}_k} \subseteq {\cal B}$. Here, ${{\cal B}_k}$ and ${\cal B}$ denote the block sets of VUE $k$'s sensory region and the whole region, respectively.
With efficient machine learning methods, the cloud server can implicitly estimate VUE $k$'s trajectory ${l_k}\left( t \right) = \left( {{x_k}\left( t \right),{y_k}\left( t \right)} \right)$ and sojourn time $\tau_{k,n}^{\rm{soj}}(t)$ with F-AP $n$ for time $t \in [0,T]$, where $ \left( {{x_k}\left( t \right),{y_k}\left( t \right)} \right)$ are coordinates.

For cooperative perception, each VUE requests a map constructing task ${\cal T}_k$ which covers its front region with a length of $d^{\text{exp}}$.
To this end, the cloud server chooses different VUEs to constitute independent cooperative clusters, and selects different block messages from cooperative VUEs for uploading. In this article, we assume that the VUEs associating with the same F-AP or the cloud server form a cooperative cluster, and partial messages could be migrated among F-APs and the cloud server via fiber backhaul links.

Once receiving these uploading messages, the cloud server and F-APs carry out customized maps for individual VUEs by sequentially realigning, integrating, inferring, and projecting the collected uploading messages. We define a tuple $(I,\mu,\tau_k^{\rm{max}})$ to refer to VUE $k$'s task, whose elements denote the data size of a block message, the number of CPU-cycle frequencies required to process 1 bit input, and ${\cal T}_k$'s maximum tolerant latency, respectively.

\subsection{Spatial-Temporal Model in Cooperative Perception}

To identify the values of block messages in cooperative perception, the spatial-temporal model is detailed as follows. 

From the perspective of timeliness, the value of VUE $k$'s block message $b$ decreases from time of generation at $t_{k,b,0}$ until it hits deadline at $t_{k,b,0}+\tau^{\text{dll}}$. A linear descend function is used to define the temporal value of VUE $k$'s block message $b$ at time $t$ as
\begin{equation}\label{eqa:t_value}
    {q_{k,b}}\left( t \right) =  - \frac{{{q_{k,b,0}}}}{\tau^{\text{dll}}}\left( {t{\rm{ - }}{t_{k,b,0}}} \right) + {q_{k,b,0}},
\end{equation}
where ${q_{k,b,0}} > 0$ is the message's initial value. 
However, considering that the sensing time $t_{k,b,0}$ and the sensing period are unknown at the cloud server, we model the temporal value $q_{k,b}\left( t \right)$ by Markov process to characterize its randomness.
For VUE $k$, the set of potential temporal value is defined as ${{\cal Q}_{k,b}} = \left\{ {{q_{k,b,1}},{q_{k,b,2}},...,{q_{k,b,Z}}} \right\}$.
${\Pr _{k,z|z'}} = \Pr \left\{ {{q_{k,b}}(t) = {q_{k,b,z}}|{q_{k,b}}({t - 1}) = {q_{k,b,z'}}} \right\}$ is the probability that the temporal value of VUE $k$'s block message $b$ at time $t$ is $q_{k,b,z}$ if its value in time $t-1$ is $q_{k,b,z'}$.

As for spatial dimension, each VUE has a rectangular region of interest where the closer sensory blocks are more expected by the VUEs.
With VUE $k$'s and block $b$'s positions $l_k(t)$, $l_b$, the euclidean distance $d_{k,b}$ between VUE $k$ and block $b$ and the angle $\theta_{k,b}$ between VUE $k$'s moving direction and block $b$  can be easily calculated, as illustrated in Fig.\ref{fig:system}.
Thus, we define the spatial value of block $b$ for VUE $k$ as 
\begin{equation}\label{eqa:s_value}
    {w_{k,b}}\left( t \right) = \frac{{d^{{\rm{exp}}} - \left| {{d_{k,b}}\left( t \right)\cos {\theta _{k,b}}\left( t \right)} \right|}}{{d^{{\rm{exp}}}}}.
\end{equation}
Finally, the spatial-temporal value of VUE $k'$'s block message $b$ for VUE $k$ can be calculated by ${u_{k,k',b}}(t) = {q_{k',b}}(t){w_{k,b}}(t)$.

\subsection{Communication and Computation Models}

As illustrated above, ${\cal T}_k$ includes block message uploading, joint processing, and output downloading process. Like\cite{du2020cooperative}, we ignore the download latency. Denote $\tau_k^{\rm{comm}}$ and $\tau_k^{\rm{comp}}$ as the uploading and computation latencies, we have ${\cal T}_k$'s sum latency $\tau_k =  \tau_k^{\rm{comm}} + \tau_k^{\rm{comp}}$.
Moreover, each task ${\cal T}_k$ could be handled by following two possible modes: the cloud mode and the F-AP mode.
Let $x_{k,n}\in \left\{ {0,1} \right\}$ denote the offloading mode selection of VUE $k$: if $x_{k,0} = 1$, VUE $k$ choose the cloud mode; while $x_{k,n} = 1, n \in \{ 1,2, ...,N\}$, VUE $k$ offloads the task to F-AP $n$.
Next, we describe the communication and computation models considered in both modes in sequence.

\subsubsection{Communication model}

In FVNETs, the total bandwidth is divided into $S$ resource blocks (RBs) with $W$ bandwidth each and full frequency reuse is considered for all F-APs and RRHs.
Define $a_{k,s}\in \left\{ {0,1} \right\}$ as the RB allocation indicator. $a_{k,s} = 1$ if RB $s$ is allocated to VUE $k$ and $a_{k,s} = 0$, otherwise.
Without loss of generality, each VUE is allocated a single RB for uploading.
For VUE $k$ associating with F-AP $n\in{\cal N}$, its uploading rate can be expressed as
\begin{equation}\label{rate_fap}
    {R_{k,n}}\left( t \right) = \sum\limits_{s \in {\cal S}} {W{{\log }_2}\left( {1 + \frac{{{a_{k,s}}{p_k}{{\left| {h_{k,n,s}^{\left( F \right)}\left( t \right)} \right|}^2}}}{{\sum\limits_{j = 1,j \ne k}^K {{a_{j,s}}{p_j}{{\left| {h_{j,n,s}^{\left( F \right)}\left( t \right)} \right|}^2}}  + {\sigma ^2}}}} \right)} ,
\end{equation}
where $p_k$, $\sigma ^2$, and ${h_{j,n,s}^{\left( F \right)}}$ are the transmit power, noise power, and channel gain from VUE $k$ to F-AP $n$ on RB $s$ at time $t$, respectively.

When VUE $k$ selects the cloud mode, it associates with its $ \left| {{\cal M}_k}  \right| $ close RRHs, whose set is described as ${\cal M}_k$.
In the meantime, optimal linear detection, i.e. minimum mean square error (MMSE) detection, is performed at the cloud server to mitigate inter-RRH interference, thus the uplink rate of VUE $k$ in the cloud mode at time $t$ is
\begin{equation}\label{eqa:rate_cloud}
    {R_{k,0}}\left( t \right) = \sum\limits_{s \in {\cal S}} {W{{\log }_2}\left( {1 + \frac{{{a_{k,s}}{p_k}{{\left| {{\bf{g}}_{k,s}^H{\bf{h}}_{k,s}^{\left( C \right)}\left( t \right)} \right|}^2}}}{{In{t_{k,s}}{\rm{ + }}{\sigma ^2}{\bf{g}}_{k,s}^H{{\bf{g}}_{k,s}}}}} \right)} ,
\end{equation}
where $Int{_{k,s}} = \sum\nolimits_{j \ne k} {{a_{j,s}}{p_j}{{\left| {{\bf{g}}_{k,s}^H{\bf{h}}_{j,k,s}^{\left( C \right)}\left( t \right)} \right|}^2}}$ is the interference and ${{{\bf{g}}_{k,s}}}$ is the MMSE detection vector.
${\bf{h}}_{k,s}^{\left( C \right)}\left( t \right) $ is the channel gain from VUE $k$ to its associated RRHs ${\cal M}_k$ on RB $s$ at time $t$,
while ${\bf{h}}_{j,k,s}^{\left( C \right)}\left( t \right)$ is the channel from VUE $j$ to the associated RRHs of VUE $k$.

Note that the offloaded task ${\cal T}_k$ starts computing only when the whole VUEs finish uploading. 
Therefore, the communication latency of task ${\cal T}_k$ relies on the maximum one among cooperative VUEs, i.e.,
\begin{equation}\label{latency_comm}
    \tau _k^{\rm{comm}}(t) = \max \left( {{{\left\{ {\frac{{\sum\nolimits_{b \in {{\cal B}_{k'}}}^{} {{e_{k',b}}I} }}{{{R_{k',n}}\left( t \right)}} + {x_{k',0}}{\tau ^{fh}}} \right\}}_{\forall k' \in {\cal K}}}} \right),
\end{equation}
where $\tau ^{fh}$ is fronthaul delay. ${e_{k,b}} \in \left\{ {0,1} \right\}$ denotes the block message selection of VUE $k$: if ${e_{k,b}} = 1$, then block message $b$ of VUE $k$ is selected for uploading.

\subsubsection{Computation model}

For cooperative perception, the offloaded task ${\cal T}_k$ has to process all uploaded block messages within its region of interest. 
Assuming that the computation capabilities of the cloud server and each F-AP $n$ are characterized by the maximum CPU-cycle frequency $f_0^{max}$ and $f_n^{max}$.
With $f_{n,k}$ CPU-cycle frequency allocated to VUE $k$, the computation latency of task ${\cal T}_k$ can be obtained by
\begin{equation}\label{latency_comp}
    \tau _{_{k}}^{comp}(t) = \frac{{\sum\limits_{k' \in {\cal K}} {\sum\limits_{b \in {{\cal T}_k}}^{} {{e_{k',b}}\mu I} } }}{{{f_{n,k}}}}.
\end{equation}

\subsection{Problem Formulation}

In FVNETs, the cloud server and F-APs are interested in finishing VUEs' offloaded tasks with more valuable block messages as quickly as possible, therefore we define the satisfaction of VUE $k$ at time $t$ as
\begin{equation}\label{eqa:satisfaction}
    {U_k}(t) = {\varepsilon _1}\sum\limits_{b \in {{\cal T}_k}} {\sum\limits_{k' \ne k}^{} {{e_{k',b}}{u_{k,k',b}}\left( t \right) + } } {\varepsilon _2}(\tau_k^{max} - {\tau _k}\left( t \right)),
\end{equation}
where $\varepsilon _1$ and $\varepsilon _2$ are weight parameters. The first part shows the overall spatial-temporal values of block messages from cooperative VUEs, and the second part denotes the effect of task latency.

The key of this work is to maximize the long-term sum satisfaction for cooperative perception in FVNETs by optimizing the offloading mode selection scheme ${\mathbf{\boldsymbol x}}^*$, the sensory block message selection scheme ${\mathbf{\boldsymbol e}}^*$, the uplink RB allocation scheme ${\mathbf{\boldsymbol a}}^*$, and the frequency resource allocation scheme ${\mathbf{\boldsymbol f}}^*$.
Mathematically,  it can be formulated as
 \begin{equation}\label{p1}
    \begin{aligned}
    \mathop {\max }\limits_{\mathbf{\boldsymbol x},\mathbf{\boldsymbol e},\mathbf{\boldsymbol a},\mathbf{\boldsymbol f}} 
    & \sum\nolimits_{t = 0}^{T {\rm{ - }}1} {\sum\nolimits_{k \in {\cal K}} {{U_k}\left( t \right)} } \\
    \mathrm{s.t.}
    & \left( {\text{a}} \right) {\tau _k}(t) \le \min \left( {\tau _k^{\max },\tau _{k,n}^{{\rm{soj}}}(t)} \right) \ \forall k\\
    & \left( {\text{b}} \right) \sum\nolimits_{k \in {\cal K}} {{e_{k,b}}}  \le 1,\ {e_{k,b}} \in \left\{ {0,1} \right\},\ \forall b\\
    & \left( {\text{c}} \right) \sum\nolimits_{n = 0}^N {{x_{k,n}}}  \le {\rm{1}},\ {x_{k,n}} \in \left\{ {0,1} \right\}, \ \forall k\\
    & \left( {\text{d}} \right) \sum\nolimits_{s \in {\cal S}} {{a_{k,s}} \le 1},\ {a_{k,s}} \in \left\{ {0,1} \right\},\ \forall k\\
    & \left( {\text{e}} \right) \sum\nolimits_{k \in {\cal K}} {{x_{k,n}}{f_{k,n}}}  \le {f^{\max}_{n}},\ \forall n \in 0 \cup {\cal N} \\
    \end{aligned}
    \end{equation}
where constraints (\ref{p1}a) shows the overall latency of individual VUE should be less than the maximum tolerant latency of ${\cal T}_k$ and its sojourn time $\tau _{k,n}^{{\rm{soj}}}$,
constraint (\ref{p1}b) regulates that the uploading block messages of different VUEs are non-overlapped,
constraints (\ref{p1}c,d) means each VUE can be allocated to a single mode and RB,
and constraint (\ref{p1}e) denotes that the allocated CPU-cycle frequencies are not allowed to exceed the frequency budgets .

\section{Joint Sensing, Communication, and Computation Resource Allocation}

Note that the formulated problem \eqref{p1} is a mixed-integer nonlinear programming problem which is in general intractable.
In this section, we decouple the original problem into a joint mode selection and block selection sub-problem plus a joint communication and computation resource allocation sub-problem. Afterwards, multi-agent DRL and swap matching-based algorithms are developed.

\subsection{Optimization of Offloading Mode and Block Selection}

Under the circumstance that communication and computation resource allocation scheme has been determined, the original problem can be reformulated as
    \begin{equation}\label{p2}
        \begin{aligned}
        \mathop {\max }\limits_{\mathbf{\boldsymbol x},\mathbf{\boldsymbol e}} 
        & \ \sum\nolimits_{t = 0}^{T {\rm{ - }}1} {\sum\nolimits_{k \in {\cal K}} {{U_k}\left( t \right)} } \\
        \mathrm{s.t.}
        & \ (\ref{p1}\text{b}),(\ref{p1}\text{c}).
        \end{aligned}
        \end{equation}
Since the block message's value is dynamic and unknown, we cannot solve problem \eqref{p2} using traditional optimization method.
Inspired by\cite{zhang2020deepreinforcementlearningbased}, we resort to multi-agent DRL to resolve this uncertainty.

\subsubsection{Multi-Agent Markov Game}
To be concrete, we treat every VUE as an intelligent agent and convert problem \eqref{p2} into a Markov game with $K$ agents.
At time $t$, each agent $k$ observes local state ${s_k(t)}$ and selects its own action ${a_k(t)}$. Given the joint actions ${\mathbf{\boldsymbol a}}(t) = (a_1(t), ... , a_K(t))$ taken by $K$ agents, the FVNET feeds back new state ${\mathbf{\boldsymbol s}}(t+1) = (s_1(t+1), ... , s_K(t+1))$ and immediate reward $r_k(t)$.
The formal definitions of three elements are given in the following.

{\bf{State:}} Each agent $k$'s local state includes the following parts:
$\tau _k^{{\rm{max}}}$, the maximum tolerant latency of VUE $k$;
${l_k}\left( t \right)$, the coordinate of its own;
${q_k}\left( {t - 1} \right)$, the temporal value of VUE $k$'s sensory messages at time $t-1$;
${\left\{ {{O_n}\left( {t - 1} \right)} \right\}_{n \in 0 \cup {\cal N}}}$, the number of serving VUEs of the cloud server and F-APs at time $t-1$;
and ${r^{sat}}\left( {t - 1} \right)$, the latency satisfied ratio.

{\bf{Action:}} Consistent with \eqref{p2}, each agent selects the offloading mode and uploading messages ${a_k}\left( t \right) = \left( {n,{{\left\{ {{e_{k,b}}} \right\}}_{b \in {{\cal B}_k}}}} \right)$.

{\bf{Reward:}} We consider this game as a fully cooperative one and define the immediate reward function of each VUE as the sum satisfaction, i.e.,
$ r_k(t) = r(t) = \sum\nolimits_{k \in {\cal K}} {{U_k}\left( t \right)}$.

\subsubsection{Attention Multi-Agent Deep Deterministic Policy Gradient
(DDPG)-based Algorithm}

Due to non-stationary local states and large action spaces, canonical DRL algorithms, like deep Q-network, are always ineffective.
Inspired by\cite{iqbal2019actorattentioncritic, sun2019applicationa}, multi-agent DDPG algorithm is tuned with an attention mechanism in this article to tackle this problem, namely attention multi-agent DDPG.

In attention multi-agent DDPG, each agent $k$ has its own actor network $\theta_k$ and critic network $\delta_k$, acting as the policy and the policy evaluator, respectively.
Note that instead of the locals, the global states and actions are considered at the critic network to address the non-stationary and cooperative issues.
Furthermore, two \emph{target networks} (${\theta_k^-}$ and ${\delta_k^-}$) and \emph{experience replay} are applied to stabilize training and to remove data correlation.
We denote $\cal D$ as the replay buffer with capacity $|\cal D|$ and express the stored experience of all agents as the tuple $\left( {\mathbf{\boldsymbol s}},{\mathbf{\boldsymbol a}},r,{\mathbf{\boldsymbol s'}} \right)$.
Here, we omit the time index $t$ and denote mark $\cdot '$ as  time $t+1$ for simplicity.

Specifically, the actor network $\theta_k$ is responsible for finding a deterministic policy $\mu_k \left( {{s_k}(t);{\theta_k }} \right)$ to maximize the cumulative discounted reward $J\left( {{\mu _k}} \right) = {{\mathds{E}}}\left[ {\sum\nolimits_{t' = t}^T {{\gamma ^{t' - t}}r\left( {t'} \right)} } \right]$, where $\gamma$ is the discounted factor.
Then, $a_k(t)$ can be obtained by
\begin{equation}\label{eq:act_select}
    {a_k}\left( t \right) = {\mu _k}\left( {{s_k}\left( t \right);{\theta _k}} \right) + {\cal N}\left( t \right),
\end{equation}
where ${\cal N}\left( t \right)$ is the stochastic noise to encourage exploration.
For its updating, the parameter $\theta_k$ is directly adjusted in the direction of ${\nabla _{{\theta _k}}}J( {{\mu _k}} )$, which is given by
\begin{equation}\label{eq:act_gradient}
    {\nabla _{{\theta _k}}}J( {{\mu _k}} ) = {{\mathds{E}}_{\mathbf{\boldsymbol s},\mathbf{\boldsymbol a} \sim {\cal D}}}\left[ {{\nabla _{{\theta _k}}}{\mu _k}\left( {{a_k}|{s_k}} \right){\nabla _{{a_k}}}{Q_k}\left( {\mathbf{\boldsymbol s},\mathbf{\boldsymbol a}} \right)} \right].
\end{equation}
Here, ${Q_k}\left( {\mathbf{\boldsymbol s},\mathbf{\boldsymbol a}} \right)$ is the action-value function that is established by the critic network $\delta_k$.
It takes as the input of the states $\mathbf{\boldsymbol s}$ and actions $\mathbf{\boldsymbol a}$ of all agents and outputs the Q-value for agent $k$.
With $\left( {\mathbf{\boldsymbol s}},{\mathbf{\boldsymbol a}},r,{\mathbf{\boldsymbol s}'} \right) \sim {\cal D}$, the weights of agent $k$'s critic network $\delta_k$ can be updated by minimizing the MSE-based loss function $L_k(\delta_k)$, i.e.,
\begin{equation}\label{eq:Q_loss}
    L\left( \delta_k  \right) = \mathds{E}_{\left( {\mathbf{\boldsymbol s}},{\mathbf{\boldsymbol a}},r,{\mathbf{\boldsymbol s}'} \right) \sim {\cal D}}\left[ {\left( y_k - { {Q_k}\left( {\mathbf{\boldsymbol s},\mathbf{\boldsymbol a}} \right) }   \right)^2} \right],
\end{equation}
where $ y_k = r + \gamma \mathop {{\rm{max}}}\limits_{a_k} Q^-  \left({\mathbf{\boldsymbol s}'},{\mathbf{\boldsymbol a}'}\right)|_{a'_k=\mu_k^-(s'_k)}$.
Here,
$Q^-  ({\mathbf{\boldsymbol s}},{\mathbf{\boldsymbol a}})$ and $\mu_k^-(s_k)$ are agent $k$'s target critic and actor networks with weights $\delta_k^-$ and $\theta_k^-$, respectively.

Furthermore, we adjust the critic network $Q_k$ with an attention mechanism which facilitates fine-grained and discriminatory treatment of different VUEs. To this end, we firstly customize a multi-layer perception (MLP) network $g_k : (s_k, a_k) \to e_k $ to reduce the input dimension ${\mathbf{\boldsymbol s}},{\mathbf{\boldsymbol a}}$ and extract the higher features ${\mathbf{\boldsymbol e}}= (e_1,...,e_K)$.
Then, agent $k$'s attention value for the other agents can be calculated by 
${v_k} = \sum\nolimits_{j \ne k} \alpha _{k,j} {{v_{k,j}}}$. Here, $\alpha _{k,j}$ and $v_{k,j}$ denote agent $k$'s attention weight and value for agent $j$, which is given by
\begin{equation}\label{eq:att_weight}
    {\alpha _{k,j}} = \frac{{\exp \left( {e_k^TW_q^T{W_k}{e_j}} \right)}}{{\sum\nolimits_{j \ne k} {\exp \left( {e_k^TW_q^T{W_k}{e_j}} \right)} }},
\end{equation}
\begin{equation}\label{eq:att_value}
    {v_{k,j}}{\rm{ = }}h\left( {{V_k}{e_j}} \right).
\end{equation}
Wherein, the parameter matrix ${W_q}$ and ${W_k}$ constitute a bilinear mapping for $e_k$ and $e_j$. In addition, $h(\cdot)$ is a ReLu function and $V$ is a transform matrix.
With the derived high feature $e_k$ and the attention value $v_k$, we could establish a new MLP network $Q(e_k,v_k)$ to replace the original critic network $Q({\mathbf{\boldsymbol s}},{\mathbf{\boldsymbol a}})$.

Finally, we conclude the attention multi-agent DDPG procedures to solving problem \eqref{p2} in \textbf{Algorithm \ref{alg1}}.

\subsection{Optimization of RB and CPU-frequency Allocation}

Once the offloading modes and selected blocks have been determined, the original problem is reduced into
\begin{equation}\label{p3}
    \begin{aligned}
    \mathop {\min }\limits_{\mathbf{\boldsymbol a},\mathbf{\boldsymbol f}} 
    & \ \sum\nolimits_{k \in {\cal K}} \tau_k \\
    \mathrm{s.t.}
    & \ (\ref{p1}\text{a}),(\ref{p1}\text{d}),(\ref{p1}\text{e}).
    \end{aligned}
    \end{equation}
Hereafter, the RBs and frequencies are successively allocated.

\begin{algorithm}[t]
    \caption{Attention multi-agent DDPG-based algorithm}
    \label{alg1}
    \begin{algorithmic}[1]
    \STATE \textbf{Initialize} actor, critic, target actor, target critic networks $\theta, \theta^-, \delta,  \delta^-$, 
     and the replay memory $\cal D$ with capacity $|{\cal D}|$.
    
    \FOR{epoch $e=1,...E$}
        \STATE Initialize the state ${\mathbf{\boldsymbol s}}$.
        \FOR{step $t=1,...T$}
            \STATE Select action $a_k$ with \eqref{eq:act_select}, $\forall k \in {\cal K}$.
            \STATE Obtain current reward $r$ and next state ${\mathbf{\boldsymbol s}}'$.
            \STATE Store tuple $\left( {\mathbf{\boldsymbol s}},{\mathbf{\boldsymbol a}},r,{\mathbf{\boldsymbol s'}} \right)$ into $\cal D$.
            \FOR{agent $k=1,...K$}
                \STATE Randomly sample a mini-batch of tuples from $\cal D$.
                \STATE Update the critic network by minimizing \eqref{eq:Q_loss}.
                \STATE Update the actor network with \eqref{eq:act_gradient}.
                \STATE Update the target critic and actor networks \\
                    ${\theta ^ - } \leftarrow \tau \theta  + \left( {1 - \tau } \right){\theta ^ - }$,
                    ${\delta ^ - } \leftarrow \tau \delta  + \left( {1 - \tau } \right){\delta ^ - }$.
            \ENDFOR
        \ENDFOR
    \ENDFOR
    \end{algorithmic}
    \end{algorithm}

\subsubsection{Swap Matching-based RB Allocation}

With fixed CPU-frequencies, we firstly model the RB allocation problem as a one-to-one matching game and solve it in a distributed and low-complexity manner.

Formally, we define $\Phi:{{\cal K}_n} \leftrightarrow {\cal S}$ as the matching function for the associated VUE set ${{\cal K}_n}, n \in 0 \cup {\cal N}$ and the bandwidth set $\cal S$, which has the following properties
\begin{equation}\label{eq:matching_pro}
    \begin{aligned}
        1) \ & k = \Phi ( s ) \ \leftrightarrow \ s = \Phi ( k ), 
        &\ \forall k \in {{\cal K}_n}, \ \forall s \in {\cal S},\\
        2) \ & \left| {\Phi(s)} \right| \le 1, \ \left| {\Phi( k)} \right| \le 1,
        &\ \forall k \in {{\cal K}_n}, \ \forall s \in {\cal S},
    \end{aligned}
\end{equation}
where condition 1) implies that if RB $s$ matches with VUE $k$, VUE $k$ also matches with RB $s$; condition 2) gives that each RB can be matched with one VUE, and each VUE can only be matched with one RB in turn.

The utility of VUE $k$ is defined as the uploading rate on RB $s$, i.e.,
\begin{equation}\label{eq:utility_vue}
    {\phi _k} = R_{k,n}^s,
\end{equation}
which indicates that every VUE selfishly favours the RBs with higher offloading rate.
As for RB $s$, it aims to minimize the overall communication latency which equals to the maximum one, thus we define the utility of RB $s$ as
\begin{equation}\label{eq:utility_rb}
    {\phi _s} = \frac{{\sum\nolimits_{b \in {B_{k'}}}^{} {{e_{k',b}}}I}}{{R_{k,n}^s}} + {x_{k',0}}{\tau ^{fh}}.
\end{equation}

Based on above analysis, a swap matching-based RB allocation algorithm is developed in \textbf{Algorithm \ref{alg2}}.
To start the matching process, each VUE $k \in {\cal K}_n$ and RB $s$ establish their own preference lists ${\cal{L}}_k$ and ${\cal{L}}_s$ in the descending order.
A deferred acceptance algorithm is then adopted for initial matching.
To overcome the dynamics introduced by inter-cell interference, two matched VUEs could exchange their matched RBs $\Phi _k^{k'} = \left\{ {\Phi {\rm{\backslash }}\left\{ {\left( {k,\Phi \left( k \right)} \right),\left( {k',\Phi \left( {k'} \right)} \right)} \right\}} \right\} \cup \left\{ {\left( {k,\Phi \left( {k'} \right)} \right),\left( {k',\Phi \left( k \right)} \right)} \right\}$
when the maximum communication latency could be reduced.

\begin{algorithm}[t]
    \caption{Swap matching-based RB allocation algorithm}
    \label{alg2}
    \begin{algorithmic}[1]
    \STATE \textbf{Initialize} the preference lists ${{\cal L}_k},{{\cal L}_s},\forall k,\forall s $ \AND
    the set of unmatched VUEs ${\cal K}^{unmatch}={\cal K}$.\\
    \WHILE{${\cal K}^{unmatch} \ne \emptyset $ \AND  $\exists {{\cal L}_k} \ne \emptyset $ } 
        \FOR{$\forall k\in{\cal K}^{unmatch}$}
            \STATE VUE $k$ proposes to its most preferred RB in ${\cal L}_k$ that has not rejected it before.
        \ENDFOR
        \FOR{$\forall s\in{\cal S}$}
            \STATE if $\sum\nolimits_{k \in {{\cal K}_n}} {{a_{k,s}} = 1} $, RB $s$ holds the matching and rejects all proposals; otherwise, RB $s$ accepts its favorite VUE and rejects the others.        
        \ENDFOR
    \ENDWHILE\\
    \WHILE{there exists a swap-pair $(k,k')$ which can reduce the maximum latency}
        \STATE Update $\Phi _k^{k'}$.
    \ENDWHILE

    \end{algorithmic}
    \end{algorithm}

\subsubsection{CPU-Cycle Frequency Allocation}

Given $\mathbf{\boldsymbol a}$, the problem \eqref{p3} is apparently a convex optimization problem with respect to $\mathbf{\boldsymbol f}$ and is easy to be solved by the interior-point method.

\section{Simulation Results and Analysis}

In this section, simulation results are provided to verify the sum satisfaction of cooperative perception in FVNETs.
We deploy $2$ RRHs, $1$ F-AP, and $6 \sim 12$ VUEs on a $1000$-meter-long road.
The whole road is equally divided into $100$ blocks and the size for each block message is $6.4$-kbits with $3$-level region quadtree.
For cooperative perception, each VUE requests a task to expand its sensory range by $150 \sim 500$ m with $\mu = 130$ and $\tau_k^{\max}=100 \sim 150$ ms.
In addition, The total bandwidth is $15$ MHz, while the cloud server and the F-AP has a maximum computation resource of $30$ GHz and $10$ GHz, respectively.

\begin{figure}[!tp]
    \centering
    \includegraphics[scale=0.55]{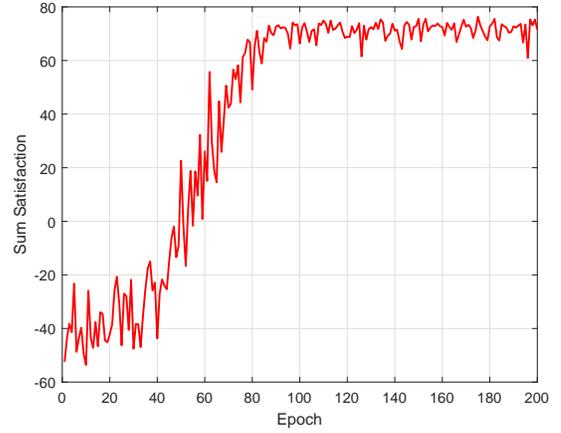}
    \caption{The convergence of algorithm \ref{alg1}.}
    \label{alg1_converge}
    \end{figure}

Fig. \ref{alg1_converge} shows the convergence performance of the proposed attention multi-agent DDPG-based algorithm, when the number of VUEs is $12$.
It is seen that our proposed algorithm could effectively converge to stable sum satisfaction.

\begin{figure}[!tp]
    \centering
    \includegraphics[scale=0.55]{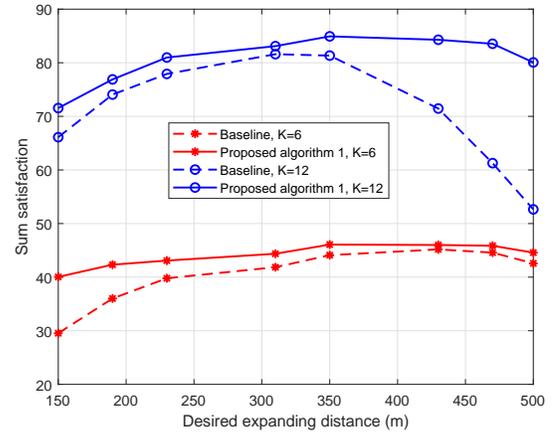}
    \caption{Sum satisfaction versus desired expanding distance.}
    \label{alg1_performance}
    \end{figure}

In Fig. \ref{alg1_performance}, the sum satisfaction performances are provided with different desired expanding distances.
For comparison, a distance-based mode selection and full message uploading scheme is adopted as the baseline.
It can be observed that our proposed attention multi-agent DDPG algorithm outperforms the baseline, because it could select more valuable block messages and avoid undesired message uploading, meanwhile balancing the load among F-APs and the cloud server. Especially when the network load is large ($K=12$), our proposed algorithm can effectively balance task load and constrained resource by controlling the selection of block messages.

\begin{figure}[!tp]
    \centering
    \includegraphics[scale=0.55]{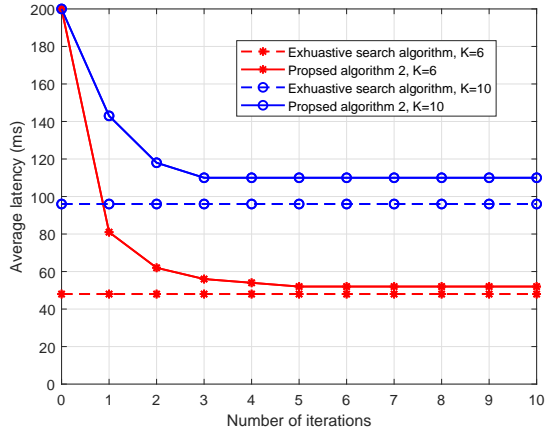}
    \caption{The convergence of Algorithm \ref{alg2}.}
    \label{alg2_converge}
    \end{figure}

Fig. \ref{alg2_converge} verifies the convergence of algorithm \ref{alg2} with different numbers of VUEs, where the optimal latency which is derived by exhaustive search is adopted for comparison.
It can be seen that, our proposed swap matching algorithm can converge to a stationary point within $3 \sim 5$ iterations, while its latency performance is close to exhaustive search algorithm.

\begin{figure}[!tp]
    \centering
    \includegraphics[scale=0.55]{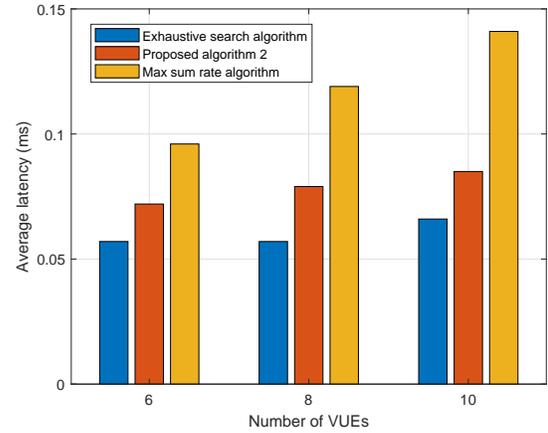}
    \caption{Average latency versus the number of VUEs.}
    \label{alg2_performance}
    \end{figure}

Fig. \ref{alg2_performance} evaluates the average latency performances with different communication RB allocation schemes, when the numbers of VUEs are set as $6 \sim 10$.
For comparison, the exhaustive search and the matching-based sum rate maximization algorithms are selected as two baselines. Wherein, the former one offers the optimal average latency.
It can be observed that our proposed algorithm outperforms the max sum rate algorithm and their latency gap becomes larger when the involved number of VUEs increases.
That is because, on the one hand, the uploading latency of cooperative perception in \eqref{latency_comm} depends on the maximum one among all cooperative VUEs rather than overall latencies of all VUEs;
on the other hand, the inter-cell interference has a larger effect on the VUE with lower channel gain.
In addition, it is also observed that our proposed algorithm achieves considerable latency performance compared with exhaustive search algorithm.

\section{Conclusion}

In this article, we focus on a vehicular cooperative perception scenario with ultra-low latency requirement, and propose a joint sensing, communication, and computation resource allocation scheme with multi-agent DRL and swap matching for
FVENTs, in which multiple VUEs constitute cooperative clusters to offload their computation tasks to either the cloud server or F-APs.
Simulation results have verified the effectiveness and superiority of our proposed algorithms on the sum satisfaction and latency performances. 
In the future, it is interesting to incorporate radio sensing and investigate corresponding resource allocation  for cooperative perception.

\section*{Acknowledgment}

This work was supported in part by the National Natural Science Foundation of China under No. 61921003 and 62001053, and the Fundamental Research Funds for the Central Universities under No. 24820202020RC11 and 2020RC03.

\bibliographystyle{IEEEtran}
\bibliography{BSTcontrol,citation_lib}

\begin{thebibliography}{10}
\providecommand{\url}[1]{#1}
\csname url@samestyle\endcsname
\providecommand{\newblock}{\relax}
\providecommand{\bibinfo}[2]{#2}
\providecommand{\BIBentrySTDinterwordspacing}{\spaceskip=0pt\relax}
\providecommand{\BIBentryALTinterwordstretchfactor}{4}
\providecommand{\BIBentryALTinterwordspacing}{\spaceskip=\fontdimen2\font plus
\BIBentryALTinterwordstretchfactor\fontdimen3\font minus
  \fontdimen4\font\relax}
\providecommand{\BIBforeignlanguage}[2]{{%
\expandafter\ifx\csname l@#1\endcsname\relax
\typeout{** WARNING: IEEEtran.bst: No hyphenation pattern has been}%
\typeout{** loaded for the language `#1'. Using the pattern for}%
\typeout{** the default language instead.}%
\else
\language=\csname l@#1\endcsname
\fi
#2}}
\providecommand{\BIBdecl}{\relax}
\BIBdecl

\bibitem{zhang2020mobile}
J.~Zhang and K.~B. Letaief, ``Mobile edge intelligence and computing for the
  internet of vehicles,'' \emph{Proc. IEEE}, vol. 108, no.~2, pp. 246--261,
  Feb. 2020.

\bibitem{yang2021machinelearningenabled}
Q.~Yang, S.~Fu, H.~Wang, and H.~Fang, ``Machine-learning-enabled cooperative
  perception for connected autonomous vehicles: Challenges and opportunities,''
  \emph{IEEE Netw.}, vol.~35, no.~3, pp. 96--101, May 2021.

\bibitem{wang2018deployment}
Y.~Wang, G.~{de Veciana}, T.~Shimizu, and H.~Lu, ``Deployment and performance
  of infrastructure to assist vehicular collaborative sensing,'' in \emph{Proc.
  {{IEEE VTC Spring}}}, {Porto, Portugal}, Jun. 2018, pp. 1--5.

\bibitem{marvasti2020cooperative}
E.~E. Marvasti \emph{et~al.}, ``Cooperative {{LIDAR}} object detection via
  feature sharing in deep networks,'' \emph{ArXiv200208440 Cs}, Feb. 2020.

\bibitem{boban2018connecteda}
M.~Boban, A.~Kousaridas, K.~Manolakis, J.~Eichinger, and W.~Xu, ``Connected
  roads of the future: Use cases, requirements, and design considerations for
  vehicle-to-everything communications,'' \emph{IEEE Veh. Technol. Mag.},
  vol.~13, no.~3, pp. 110--123, Sep. 2018.

\bibitem{higuchi2019valueanticipating}
T.~Higuchi, M.~Giordani, A.~Zanella, M.~Zorzi, and O.~Altintas,
  ``Value-anticipating {{V2V}} communications for cooperative perception,'' in
  \emph{Proc. {{IEEE IV}}}, {Paris, France}, Jun. 2019, pp. 1947--1952.

\bibitem{abdel-aziz2020vehicular}
M.~K. {Abdel-Aziz}, C.~Perfecto, S.~Samarakoon, M.~Bennis, and W.~Saad,
  ``Vehicular cooperative perception through action branching and federated
  reinforcement learning,'' \emph{ArXiv201203414 Cs}, Dec. 2020.

\bibitem{du2020cooperative}
H.~Du, S.~Leng, K.~Zhang, and L.~Zhou, ``Cooperative sensing and task
  offloading for autonomous platoons,'' in \emph{Proc. {{IEEE GLOBECOM}}},
  {Taipei, Taiwan}, Dec. 2020, pp. 1--6.

\bibitem{zhang2021delayoptimized}
K.~Zhang, M.~Peng, and Y.~Sun, ``Delay-optimized resource allocation in
  fog-based vehicular networks,'' \emph{IEEE Internet Things J.}, vol.~8,
  no.~3, pp. 1347--1357, Feb. 2021.

\bibitem{zhang2020deepreinforcementlearningbased}
X.~Zhang, M.~Peng, S.~Yan, and Y.~Sun, ``Deep-reinforcement-learning-based mode
  selection and resource allocation for cellular {{V2X}} communications,''
  \emph{IEEE Internet Things J.}, vol.~7, no.~7, pp. 6380--6391, Jul. 2020.

\bibitem{iqbal2019actorattentioncritic}
S.~Iqbal and F.~Sha, ``Actor-attention-critic for multi-agent reinforcement
  learning,'' in \emph{Proc. {{ICML}}}, {California, USA}, May 2019, pp.
  2961--2970.

\bibitem{sun2019applicationa}
Y.~Sun, M.~Peng, Y.~Zhou, Y.~Huang, and S.~Mao, ``Application of {{Machine
  Learning}} in {{Wireless Networks}}: {{Key Techniques}} and {{Open
  Issues}},'' \emph{IEEE Commun. Surv. Tutor.}, vol.~21, no.~4, pp. 3072--3108,
  2019.

\end{thebibliography}

\end{document}